\documentclass[
reprint,
superscriptaddress,
showpacs,
,preprintnumbers,
 amsmath,amssymb,
 aps,
 prx
]{revtex4-1}

\usepackage[colorinlistoftodos]{todonotes} 
\usepackage{graphicx}
\usepackage{color}
\usepackage{dcolumn}
\usepackage{bm}
\usepackage{amssymb}   
\usepackage{hyperref}

\begin{document}

\title{Bright $\gamma$ rays source and nonlinear Breit-Wheeler pairs in the collision of high density particle beams}

\author{F. Del Gaudio}
\email{fabrizio.gaudio@tecnico.ulisboa.pt}
\affiliation{GoLP/Instituto de Plasmas e Fus\~ao Nuclear, Instituto Superior T\'ecnico, Universidade de Lisboa, 1049-001 Lisbon, Portugal}

\author{T. Grismayer}
\email{thomas.grismayer@ist.utl.pt}
\affiliation{GoLP/Instituto de Plasmas e Fus\~ao Nuclear, Instituto Superior T\'ecnico, Universidade de Lisboa, 1049-001 Lisbon, Portugal}

\author{R. A. Fonseca}
\affiliation{GoLP/Instituto de Plasmas e Fus\~ao Nuclear, Instituto Superior T\'ecnico, Universidade de Lisboa, 1049-001 Lisbon, Portugal}
\affiliation{DCTI/ISCTE Instituto Universit\'ario de Lisboa, 1649-026 Lisboa, Portugal}

\author{W. B. Mori}
\affiliation{Departments of Physics \& Astronomy and of Electrical Engineering, University of California Los Angeles, Los Angeles, CA 90095, United States of America}

\author{L. O. Silva }
\email{luis.silva@ist.utl.pt}
\affiliation{GoLP/Instituto de Plasmas e Fus\~ao Nuclear, Instituto Superior T\'ecnico, Universidade de Lisboa, 1049-001 Lisbon, Portugal}

\date{\today}

\begin{abstract}

The collision of ultrashort high-density $e^-$ or $e^-$ and $e^+$ beams at 10s of GeV, to be available at the FACET II and in laser wakefield accelerator experiments, can produce highly collimated $\gamma$ rays (few GeVs) with peak brilliance of $10^{27}~\mathrm{ph/s~mm^2 mrad^2 0.1\% BW}$ and up to $10^5$ nonlinear Breit-Wheeler pairs. We provide analytical estimates of the photon source properties and of the yield of secondary pairs, finding excellent agreement with full-scale 3D self-consistent particle-in-cell simulations that include quantum electrodynamics effects. Our results show that beam-beam collisions can be exploited as secondary sources of $\gamma$ rays and provide an alternative to beam-laser setups to probe quantum electrodynamics effects at the Schwinger limit.

\end{abstract}


\maketitle
\section{Introduction}
Colliders are a cornerstone of fundamental physics of paramount importance to probe the constituents of matter.
At the interaction point of a collider, several detrimental beam-beam effects should be avoided, chief among these are beam disruption~\cite{Hollebeek1981NIM,Chen1988PRD}, beamstrahlung radiation~\cite{Chen1992PRD,Blankenbecler1987PRD}, and pair creation~\cite{Chen1989PRL}.
Beam disruption arises when the collective field of these beams focuses (unlike charges $e^-e^+$),  deflects, or blows apart (like charges $e^-e^-$) each beam~\cite{Hollebeek1981NIM,Chen1988PRD} such that the beam density profile is significantly altered or the number of collisions may be reduced.
During this process, beamstrahlung photons are emitted via nonlinear Compton scattering, and in turn can decay into electron-positron pairs via the multi-photon Breit-Wheeler mechanism~\cite{Chen1989PRL,Breit1934PR}.
They become more relevant in the quantum regime, when the relativistic invariant parameter $\chi$ exceeds unity.
The parameter $\chi=\frac{1}{E_s}\sqrt{(\gamma {\bf E} + \frac{{\bf p}\times {\bf B}}{mc})^2 - (\frac{{\bf p} \cdot {\bf E}}{mc})^2}$~\cite{Ritus1985JSLR} measures the closeness to the Schwinger limit $E_s = m^2c^3/e\hbar$ of a particle with momentum ${\bf p}$ and Lorentz factor $\gamma$, crossing an electromagnetic field ${\bf E},~{\bf B}$, where $m$ is the electron rest mass, $c$ is the speed of light, $e$ is the elementary charge and $\hbar$ is the Planck constant.
The classical regime is identified by $\chi\ll1$, the full quantum regime by $\chi\gg1$ and the quantum transition regime ranges from $0.1 \lesssim \chi \lesssim 1$.
In designs for linear colliders at the energy frontier (TeVs) disruption and beamstrahlung can be major issues due the large charge and small spot sizes that are needed to achieve large luminosity.
As a result, such linear colliders are usually designed to avoid beam disruption and the quantum regime by using flat and elongated beams \cite{TecRevLC1995} since photons and secondary pairs represent an energy loss for the beams~\cite{Noble1987NIMA,Blankenbecler1988PRL,ChattopadhyayPROC1996} and a source of background noise for the detectors~\cite{Chen1989PRL,ChattopadhyayPROC1996}, respectively.
{\color{black} On the contrary, in this work we show that  disruption, beamstrahlung, and pair production, which were previously regarded as detrimental effects, do have appealing potential from a fundamental physics and from a secondary source perspective.}
The quantum regime is actually accessible in electron-positron, or similarly electron-electron, collisions of round  beams at considerably lower energy ($\mathcal{E}\sim 10s~\mathrm{GeV}$), if the spot size at the collision is small ($\sigma_0\sim \mathrm{\mu m}$) and the beam current  ($I\sim 100s~\mathrm{kA}$) is high. 
Such beams should be available at the new SLAC facility, FACET II \citep{White2017FACET}, and in the next generation of Laser Wakefield Accelerators (LWFA) experiments, opening new exciting opportunities.
LWFA experiments already deliver beams with  10s of kAs of current and micron spot sizes within a single acceleration stage~\cite{Kneip2009PRL,Hafz2009NP,Leemans2006NP}, and are advancing towards a multistage configuration to reach higher energies~\citep{Steinke2016NAT,Kim2013PRL}, with the ultimate goal of a TeV LWFA collider~\citep{Leemans2009PT}.
Recent advances, along with theoretical models and full scale simulations \cite{LuPRSTAB2007,Silva2009CRP,MartinsNP2010}, predict 10-30 GeV beams from LWFA accelerators driven by 250 J class lasers in a single stage that should be soon available \cite{eli,vulcan,apollon}.

In this Article, we show that bright $\gamma$ rays and copious secondary pairs are produced during the collision of electron-electron and electron-positron beams for the range of parameters soon to be available.
We introduce an analytical model that is in excellent agreement with 3D QED-PIC simulations performed with QED-OSIRIS~\cite{OSIRIS,OSIRIS2,Grismayer2017PRE,Grismayer2016PoP,Vranic2016NPJ,Vranic2017PPCF}.
Our results attest that, at the threshold of the quantum regime, beamstrahlung and pair production are driven by the maximum collective field region, with a clear experimental signature, which cannot be described by the uniform average field model~\citep{Chen1992PRD,Chen1989PRL}.
As a consequence, the yield of secondary pairs can be orders of magnitude higher than what has been predicted before~\citep{Chen1989PRL}.
This can open the door to the experimental observation of nonlinar Breit-Wheeler pair production.
Furthermore, the collisions of ultra high current electron beams at 10's of GeV could be exploitable as a secondary source of GeV $\gamma$ rays reaching peak brilliance of $10^{27}~\mathrm{ph/s~mm^2 mrad^2 0.1\% BW}$ leveraging on the small divergence angle at which the $\gamma$ rays are emitted during the beam-beam interaction.
This idea was briefly mentioned but overlooked in the context of TeV colliders \citep{Blankenbecler1988PRL,Chen1992AIP}.

\section{Physical picture}
The propagation of a single relativistic beam is almost force-free as the contribution of the beam space charge and the beam current to the Lorentz force nearly balance to the order of $1/\gamma^2$~\cite{KatsouleasPFB1990}.
However, when two counter-propagating beams overlap, either the charge density ($e^-e^+$ collisions) or the current density ($e^-e^-$collisions) vanishes, such that the two contributions to the Lorentz force are no longer in balance.
This generates very large focusing ($e^-e^+$) or defocussing ($e^-e^-$) forces which lead to the disruption of the beams.
The magnitude of this effect is measured through the disruption parameter ~\cite{Hollebeek1981NIM}
\begin{equation}
D=\sqrt{\pi/2}~(\omega_b^2/\gamma) (\sigma_z/c)^2,
\end{equation}
where $\omega_b=\sqrt{4\pi e^2 n_0/m}$ is the plasma frequency associated with the beam peak density $n_0$, and $\sigma_z$ is the beam length.
This parameter is related to the number of relativistic plasma periods contained within the beam duration time ($\sim\sqrt{D}$) and it distinguishes three regimes: i) the low disruption regime $D<1$, ii) the transition regime $1< D < 10$, and iii) the confinement regime $D>10$.
The disruption parameter can also be cast in the form
\begin{equation}
D\simeq 0.075 (\sigma_z/\sigma_0)^2 I[\textrm{MA}]/\mathcal{E}[\textrm{GeV}]\sim 10^{-3},
\end{equation}
for the beam parameters discussed above.
In the low disruption regime, the beam duration (collision time) is considerably smaller than the relativistic plasma period and the particles are almost free streaming leading to similar interaction dynamics for $e^-e^-$ and $e^-e^+$ collisions.
Even for low disruption, beamstrahlung radiation and consequent pair production can rise significantly if the two beams interact in the quantum regime.
Beamstrahlung and pair production are commonly described via the parameter $\Upsilon_{\mathrm{mean}}$ which is the ratio of an effective mean field strength over the Schwinger field~\citep{Chen1992PRD,Chen1989PRL}.
However, this description fails to describe the quantum transition regime where photon emission and pair production are driven by the maximum field region.
For round colliding beams, the maximum $\chi$ is 
\begin{equation}
\hat{\chi}\simeq 0.081~I\mathrm{[MA]}~\mathcal{E}\mathrm{[GeV]} / \sigma_0\mathrm{[\mu m] }\simeq 0.1-1,
\end{equation} for the beam parameters discussed above.
In particular, the transition from the classical to the quantum regime is marked by an exponential growth of the Breit-Wheeler cross section with respect to the local value of $\chi$.
In this regime the detailed description of the field configuration is of absolute importance.
{\color{black} The field of a beam resembles a half cycle laser with wavelength $\lambda\simeq 4\sigma_z$.
The field nonlinearity parameter of a laser is $a_0=eE/mc\omega_0$, where $\omega_0$ is the laser frequency.
Similarly, the parameter $a_0$ can be estimated for the collective field of a relativistic beam as $a_0= \sqrt{\frac{2}{\pi^3}}\frac{r_e}{\sigma_0}N$, where $N$ is the number of particles in the beam and $r_e$ is the classical electron radius.
For $a_0\ll1$ the pair production process involves two photons, one emitted by beamstrahlung and one from the collective electromagnetic field.
For $a_0\gg1$  the pair production process becomes multi-quantum (non-linear Breit-Wheeler), it involves one photon emitted by beamstrahlung and several photons from the collective field.
The nonlinear Breit-Wheeler process has only been approached experimentally \cite{BurkePRL1997}. 
}

\section{Analytical model}
In the proper frame of reference (primed) of one beam, the collective field is purely electrostatic ${\bf E}^{'}\neq0,~{\bf B}^{'}=0$.
In the laboratory frame (unprimed), this field is Lorentz transformed as a crossed electromagnetic field, ${\bf E}=\gamma{\bf E}^{'}-\frac{\gamma^2}{\gamma+1} {\bm \beta}( {\bm \beta} \cdot {\bf E}^{'} )$ and ${\bf B}=\gamma{\bm \beta}\times{\bf E}^{'}$ where ${\bf B}_{\perp}$ is perpendicular to ${\bf E}_{\perp}$ with the same magnitude $B_{\perp}=E_{\perp}\sqrt{1-1/\gamma^2}\simeq E_{\perp}$ and $E_{\parallel}= E_{\perp}/\gamma$, $B_{\parallel}=0$, with the parallel component along $z$, the propagation axis, and the perpendicular component along $r$.
Particles are deflected by the collective fields giving rise to the disruption of the beams and to a growth in the emittance $\epsilon \simeq D \sigma_0^2/2\sigma_z$.
However in the low disruption regime, particles are almost free streaming, and the field configuration is not altered during the two beams interaction time.
Thus, the local parameter $\chi$ is given by $\chi=2\gamma E_{\perp}/E_s$.
If either the beam energy loss is negligible or the number of emitted photons is less than one per primary particle, then $\gamma$ can be assumed to be constant.
In this case, the local value of the differential probability rate of photon emission $W_{\omega}$ and the rate of pair production $W_p$ are determined uniquely by the local value of $\chi(E_{\perp})$.
The photon spectrum generated by a primary electron that crosses such field is given by $s_{\omega}(\xi,r)=\int_{-\infty}^{\infty} W_{\omega}dt$ where $t=z/2c$ and $\xi = \hbar\omega/\gamma mc^2$ is the normalized photon energy.
On its turn, a photon of energy $\xi$ emitted at time $t$ has a probability $\mathcal{P}_{\omega\rightarrow p}(r,t)=\int_t^{\infty}W_p dt'$ to decay in an electron-positron pair.
The joint probability to produce a secondary pair from a primary particle is then $y_p(r) = \int_0^1 \int_{-\infty}^{\infty} W_{\omega}\mathcal{P}_{\omega\rightarrow p} dt d\xi$.
We consider the field $E_{\perp}=4\pi en_0\sigma_0\exp(-z^2)\left[ 1-\exp(-r^2)\right]/(\sqrt{2}r)=4\pi en_0\sigma_0 f(r,z)$ of a beam with Gaussian density profile, $n(r,z)= n_0~\exp{\left(-r^2-z^2\right)}$, where $r$ and $z$ coordinates are normalized to $\sqrt{2}\sigma_0$ and to $\sqrt{2}\sigma_z$ respectively.
By averaging $s_{\omega}(\xi,r)$ over the beam density profile, we obtain the collective photon spectrum:
\begin{subequations}
\begin{align}
\mathcal{S}_{\omega}(\xi) & =  \frac{\int_0^{\infty} n(r)s_{\omega}\left(\xi,r\right) rdr}{ \int_0^{\infty} n(r) rdr}  \label{eq: phspec} \\
  & \simeq  \frac{\alpha^2\sigma_z}{2\sqrt{3}r_e\gamma} \left[ \frac{9}{5} \sqrt{\frac{2}{\pi}} \Xi_{2/3} + \frac{\xi^2}{1-\xi} \Xi_{1/2}  \right] \label{eq: phspecapp}
  \end{align}
\end{subequations}
with
\begin{eqnarray}
\Xi_{\nu}  &=&  \frac{l_{\nu}\exp{(-m_{\nu})}}{1+m_{\nu}} \left\lbrace 1+\sqrt{\pi} \Sigma_{\nu} \left[ 1+\mathrm{erf}\left( \Sigma_{\nu}\right) \right] \exp{\left( \Sigma_{\nu}^2\right)} \right\rbrace, \nonumber \\
\Sigma_{\nu}&=&\frac{m_{\nu}}{\sqrt{1+m_{\nu}}}, \nonumber
\end{eqnarray}
where $l_{\nu} = \frac{\hat{b}^{-\nu}}{\sqrt{\nu+\hat{b}}}\exp{(-\hat{b})}$ and $m_{\nu} = \frac{3}{5}\left[\hat{b}+\nu+\frac{\hat{b}}{2(\nu+\hat{b})}\right]$ (For a derivation of $l_{\nu}$ and $m_{\nu}$ see Appendix~\ref{App: Photon}).
In the classical regime the parameter $b=\frac{2}{3\chi}\frac{\xi}{1-\xi}$ reduces to $b\simeq\frac{2}{3\chi}\xi\simeq \frac{\xi}{\xi_c}$ where $\xi_c=3\chi/2=\hbar\omega_c/\gamma mc^2$ is the critical frequency for synchrotron radiation, $\omega_c\simeq 3\gamma^2 e (2B_{\perp})/2mc$.
Here $\hat{b}  =  \frac{2}{3\hat{\chi}}\frac{\xi}{1-\xi} $ refers to the maximum field region located around $r=1,~z=0$.
The fraction of radiated photons is $\mathcal{Y}_{\omega}=\int_0^1\mathcal{S}(\xi) d\xi$ producing in the beam a fractional average energy loss of $\eta_{\omega}=\int_0^1\xi \mathcal{S}(\xi) d\xi$ where the mean photon energy is $\bar{\xi}=\eta_{\omega}/\mathcal{Y}_{\omega}$.
The radiated photons are confined within the portion of space occupied by the beam, thus the peak photon flux is $\Phi_{\omega}\simeq\mathcal{Y}_{\omega} N c / \sqrt{2\pi}\sigma_z$.
If the disruption angle $\alpha_D\sim r_eN/\sigma_0\gamma$ is larger that the photon emission cone $\alpha_{\omega}\sim1/\gamma$, the divergence of the photon source is driven by the emittance of the primary beam and the photon brightness amounts to $ \Phi_{\omega}/\epsilon^2$.
At the threshold of the quantum regime, the majority of beamstrahlung energy is emitted and converted into pairs around the region of maximum field ($\hat{\chi}=8\pi \frac{r_e^2}{\alpha}\gamma n_0 \sigma_0 f(1,0)$).
By averaging $y_p(r)$ over the beam density profile, we obtain the yield of the secondary pairs:
\begin{eqnarray}
\mathcal{Y}_p & = & \frac{\int_0^{\infty} n(r)y_p(r) rdr}{ \int_0^{\infty} n(r) rdr}  \nonumber\\
 & \simeq & \frac{1}{\sqrt{6}}\left(\frac{\pi\alpha^2\sigma_z}{15r_e\gamma}\right)^2 \left[ \frac{9}{5} \sqrt{\frac{2}{\pi}} \Xi_{2/3} + \frac{4}{3} \Xi_{1/2}  \right]. \label{eq: pairs}
\end{eqnarray}
Here, the function $\Xi_{\nu}$ applies to $l_{\nu} = \frac{\hat{\psi}^{-(3/2+\nu)}\exp{\left(-4\hat{\psi}\right)}}{\sqrt{(\nu+\hat{\psi})(1+3\hat{\psi})}}$ and $m_{\nu} =\frac{3}{5} \left[4\hat{\psi}+\frac{3}{2}+\nu + \frac{\hat{\psi}}{2}\frac{(1+3\hat{\psi})+3(\nu+\hat{\psi})}{(\nu+\hat{\psi})(1+3\hat{\psi})} \right] $, where $\hat{\psi}=4/3\hat{\chi}$ (Full derivation in Appendix~\ref{App: Pairs}).
This model works within the following limits: $D\ll 1$, $\hat{\chi}\lesssim1$, and $\eta_{\omega}\ll1$ or $\mathcal{Y}_{\omega}\lesssim1$.

\section{Simulations}
To illustrate and to complement our analytical estimates, we simulated the collision of two beams for a wide range of conditions. Here we consider beams with $\sigma_0=1$ $\mathrm{\mu m}$, $\sigma_z=3$ $\mathrm{\mu m}$, the number of particles in each beam is $N=4,~5,~5.8\times10^{10}$, for energies of $\mathcal{E}=25,~30$ GeV.
The center of each beam is set at $z_c=\pm3\sigma_z$ at $t=0$, and they counter propagate for a total simulated time of $6~\sigma_z/c$, when the two beams completely crossed each other.
Each beam is composed of $5\times10^7$ simulation particles.
The simulation box size is $L_z \gtrsim 18\sigma_z$, $L_x=L_y\gtrsim 6\sigma_0$.
The spatial and  temporal resolutions range from $dx=dy=0.1$ to $dz=0.4~c/ \omega_b$ and $dt=0.007-0.07~\omega_b^{-1}$.
The collective field corresponding to the initial beam density in vacuum is initialized with a Poisson solver~\cite{Balzarini2015EPS} in the proper frame of reference of each beam and then Lorentz transformed into the laboratory frame. 
This allows a self-consistent treatment of the initial field configuration at the interaction point.
The photon emission rate in the simulations does respect the local constant field approximation~\cite{DiPiazzaPRA2018} (See Appendix~\ref{App: LCFA} for a detailed discussion).

\begin{figure}[tb]
\includegraphics[width=8.6 cm]{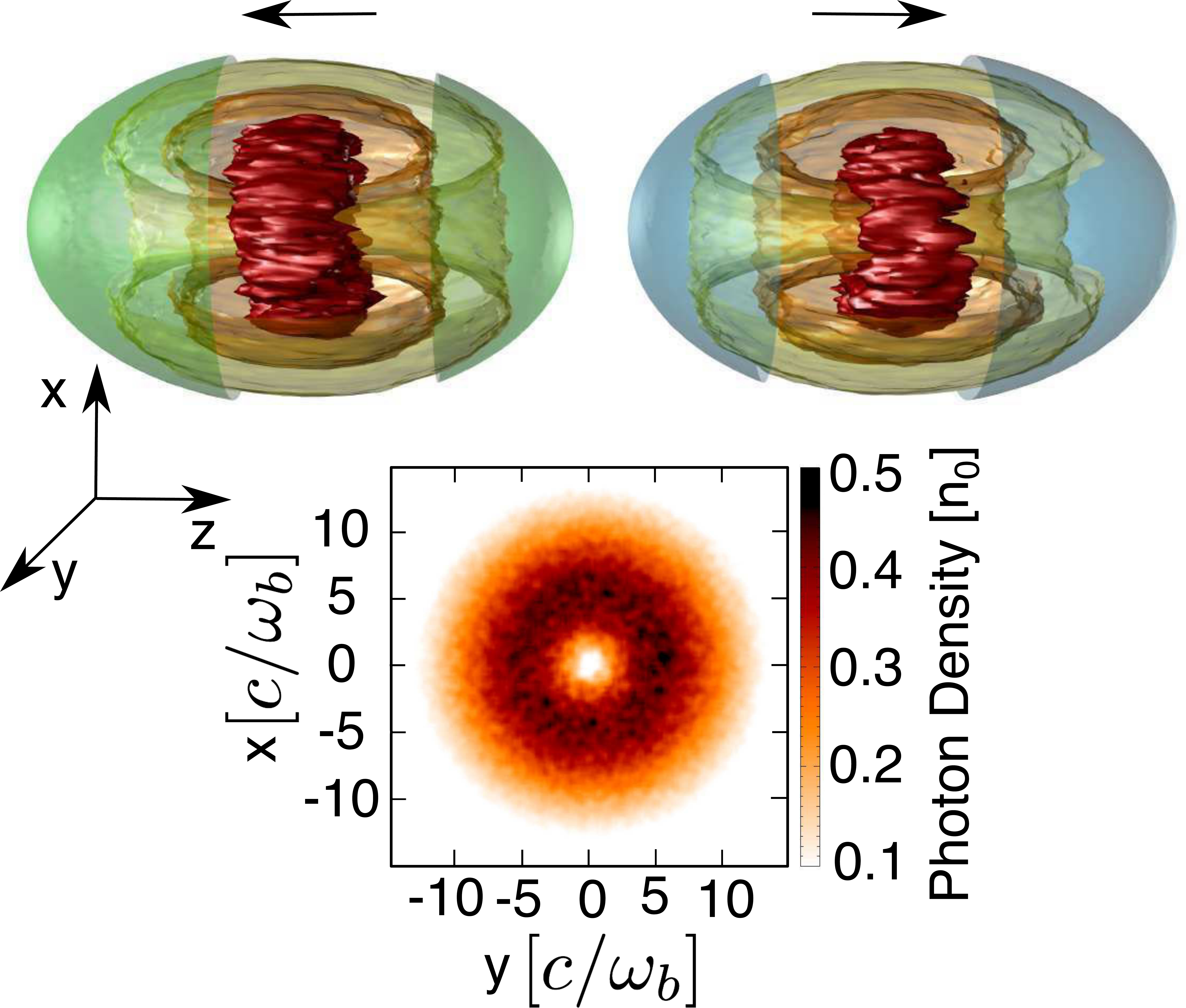}
\caption{The produced photon beam is represented by different iso-surfaces of the photon density [$n_{\omega}=0.35,~0.5,~0.65~(n_0)$] (yellow, orange, and red, respectively). The photon density is zero on axis, where the collective field vanishes, and it is maximum for $R\sim5~\mathrm{c/\omega_b}$. The positron beam is shown in green, and the electron beam in blue.}
\label{fig: image}
\end{figure}

The produced photon density (seen on Fig.~\ref{fig: image}) vanishes on axis, where the collective field vanishes and photon emission and pair production are suppressed.
On the other hand the probability of photon emission is maximum around $r \simeq1$, where the collective field is maximum.
However, it is the joint contribution of both the collective field and the density shape of the beam that determines the radius where the photon density is maximum.
On a plane perpendicular to the photon propagation direction, the photon beam has a ring shape which reveals that beamstrahlung occurs mostly in a specific region of space.
In the peak field region, the regime $\chi\sim1$ is approached.
Here, the likelihood of emitting energetic photons as well as the likelihood for these photons to decay into new secondary pairs increases exponentially.

\begin{figure}[h!]
\includegraphics[width=8.6 cm]{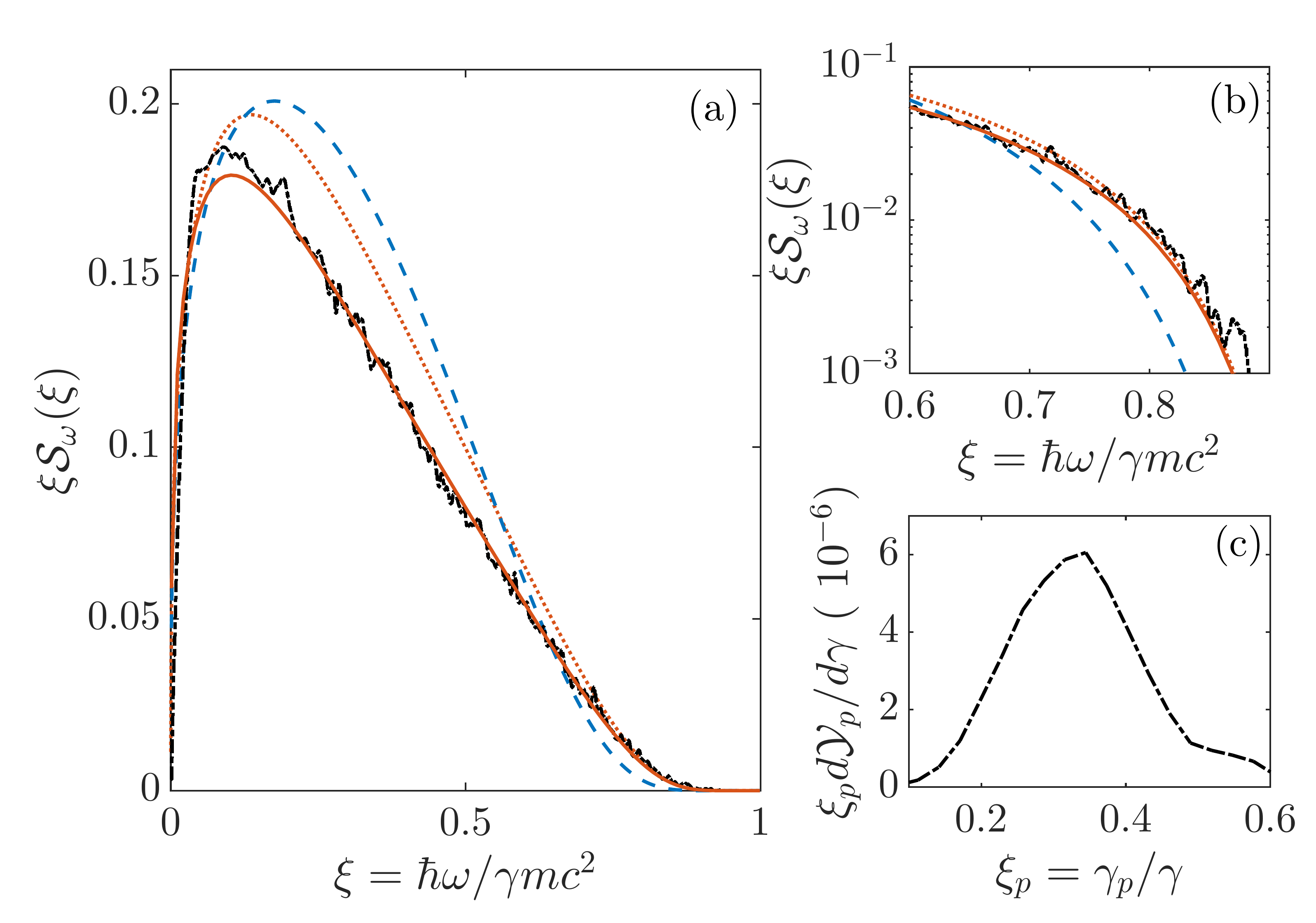}
\caption{ (a, b) Beamstrahlung energy spectrum, for two colliding beams of $N=5.8\times10^{10}$ particles, $\mathcal{E}=30$ GeV energy, $\sigma_0=1~\mathrm{\mu m}$ spot size and $\sigma_z=3~\mathrm{\mu m}$ length. The blue dashed line represents the uniform averaged field model \citep{Chen1992PRD}, the black dot-dashed line represents the OSIRIS simulation, the red solid line is the numerical evaluation of Eq.(\ref{eq: phspec}), and the second red dotted line is the analytical approximation of Eq.(\ref{eq: phspec}) given by Eq.(\ref{eq: phspecapp}). (c) Energy spectrum of secondary pairs energy from the OSIRIS simulation.}
\label{fig: Spectrum}
\end{figure}

Figure~\ref{fig: Spectrum} (a, b) show the beamstrahlung photon energy spectrum predicted by Eq.(\ref{eq: phspec}), Eq.(\ref{eq: phspecapp}) and obtained from the simulations in comparison with the uniform average field model~\citep{Chen1992PRD}.
As it can be seen in Fig.~\ref{fig: Spectrum} (b), the uniform average field model underestimates the number of beamstrahlung photons in the high energy tail which are the photons with the highest probability to decay into secondary pairs.
Figure~\ref{fig: Spectrum} (c) shows the secondary pairs energy spectrum, with a very large mean energy of $\bar{\mathcal{E}_p} \sim 10 ~ \mathrm{GeV}$, and a small energy spread (rms) $\Delta \mathcal{E}_p /\bar{\mathcal{E}_p}\sim$30\%, as compared with other secondary sources of neutral lepton jets \cite{ChenPRL2015,Sarri2015NC}.

\begin{figure}[tb]
\includegraphics[width=8.6 cm]{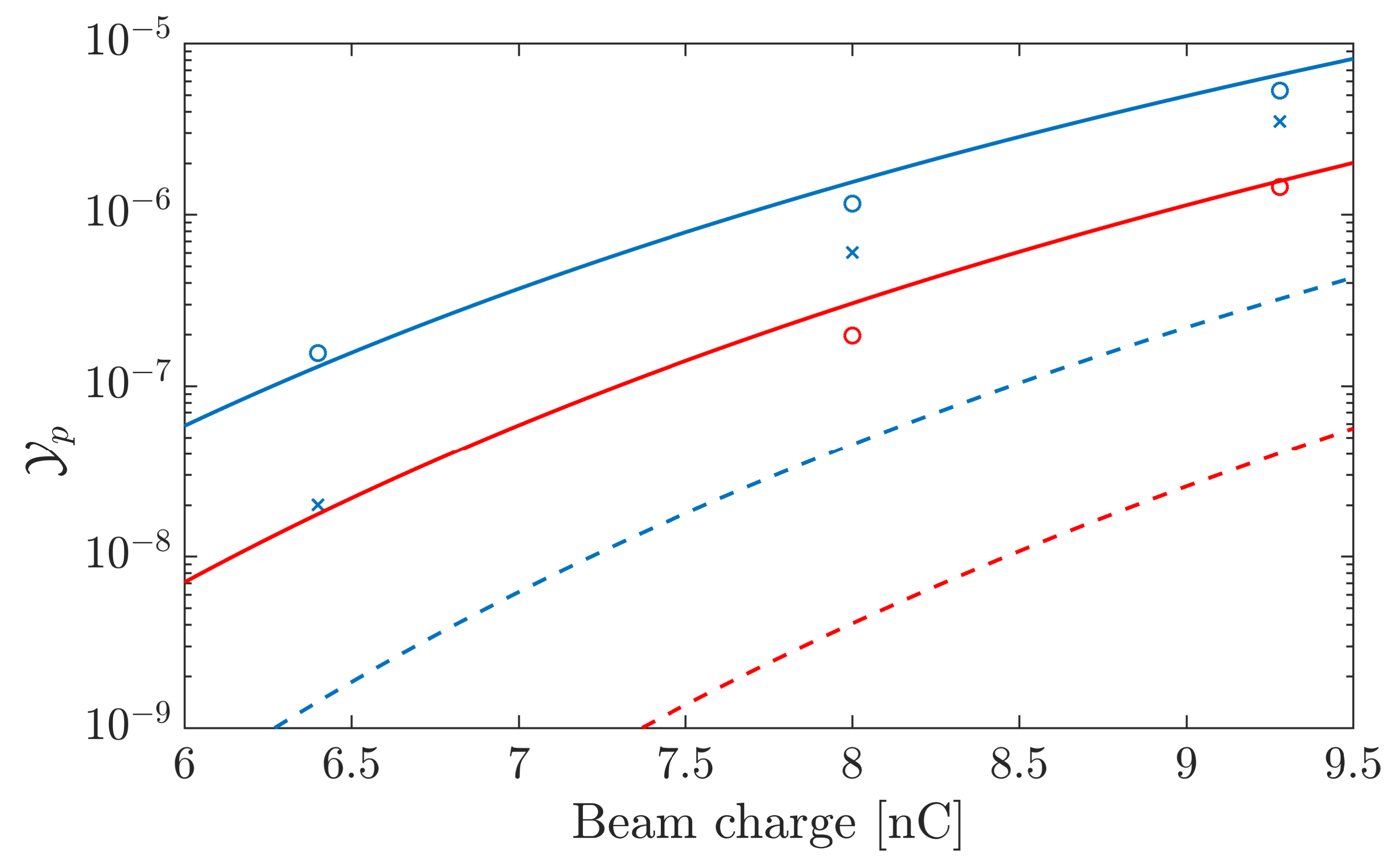}
\caption{Yield of the secondary pairs for two colliding beams with $\sigma_0=1~\mathrm{\mu m}$ spot size and $\sigma_z=3~\mathrm{\mu m}$ length, for $\mathcal{E}=25,~30$ GeV energy (red and blue, respectively), with the dashed line for the uniform averaged field model \citep{Chen1989PRL}, OSIRIS simulation results represented by ($\circ$), and the solid line representing Eq.(\ref{eq: pairs}). The uniform average field model underestimates significantly the pair yield. The OSIRIS simulations for elliptic beams ($\sigma_x/\sigma_y=2.25$) are denoted by ($\times$).}
\label{fig: 3D_Pairs}
\end{figure}

Figure~\ref{fig: 3D_Pairs} shows the yield of secondary pairs predicted by Eq.(\ref{eq: pairs}) and obtained from simulations in comparison with the uniform average field model \cite{Chen1989PRL}.
As the pair production process is exponentially dependent on the local field strength, only the highest region of the collective field drives secondary pair production.
With this regard, the uniform average field model underestimates the yield of secondary pairs.
We performed as well simulations with elliptic beams, $\sigma_x/\sigma_y=2.25$, where the transverse beam section is preserved ($\sigma_0^2=\sigma_x\sigma_y$).
We observe a reduced pair yield for the elliptic beams compared to the round ones in agreement with the qualitative predictions~\citep{Chen1989PRL}.

\section{Discussion and conclusions}
In order to contestualize the properties of the source predicted for this configuration, we report, in Table~\ref{tab comparison sm}, the characteristics of the beamstrahlung $\gamma$ rays source compared to existing and proposed sources~\cite{RobinsonNJP2010,GonoskovPRX2017,ChenPRL2013,ThomasPRX2012}.
The beamstrahlung source dimensions at the interaction point are determined by the primary beam both in spot size and duration.
In the case considered here, the transverse spot size is $1~\mathrm{\mu m}$ and the duration is $10$ fs.
The high brightness $10^{30}$ $\mathrm{ph/s~mm^2 mrad^2}$ and brilliance $10^{27}$ $\mathrm{ph/s~mm^2 mrad^2 0.1\% BW}$ are achieved operating in the low disruption regime that allows to confine the radiation over a small angle $\alpha_{D}\simeq 2.7\times 10^{-3}$ rad.

\begin{table}[b!]
\caption{Beamstrahlung $\gamma$ rays source, simulation (BS) and model (BM) for beams of $N=5.8\times10^{10}$, $\mathcal{E}=30$ GeV, $\sigma_0=1$ $\mu m$ and $\sigma_z=3$ $\mu m$. Comparison with free electron lasers (FEL)~\cite{RobinsonNJP2010}, synchrotron sources (SY)~\cite{RobinsonNJP2010},  exotic laser-based (ELB) sources~\cite{GonoskovPRX2017}, Compton sources (LCS) \cite{ChenPRL2013}, non-linear Compton sources (NLCS) \cite{ThomasPRX2012}. Photon flux $\Phi_{\omega}$ in $\mathrm{ph/s}$, peak brilliance in $\mathrm{ph/s~mm^2 mrad^2 0.1\% BW}$, and energy in eV.}
\label{tab comparison sm}
\begin{ruledtabular}
\begin{tabular}{cccccc}
		 &  $\eta_{\omega}$      	&  $ \mathcal{Y}_{\omega}$   & $\Phi_{\omega}$	& Brilliance  & Energy \\
\hline 
BS 		 &$0.085$		 		&  $0.95$				  & $2.2\times10^{24}$ 	&    $9\times10^{26}$     &      $10^{9}$ \\
BM		 & $0.096$     			&  $0.96$    			& $1.9\times10^{24} $  	&   $4.5\times10^{26} $  &	-	\\
FEL		 & - 					& - 					& - 					& $10^{28}-10^{34}$ &    $10- 10^{5}$\\
SY		 & - 					& - 					& - 					& $10^{19}-10^{25}$ &     $1-10^{6}$ \\
ELB           & $10^{-3}-0.4$		& - 					& - 					& $10^{23}-10^{26}$ &     $10^{7}-10^9$ \\
LCS		 &-					&- 					&- 					& $10^{19}$ 		& $10^6$ \\
NLCS             & 0.02 				&-					&-					& $10^{29}$ 		&     $10^7$ \\
\end{tabular}
\end{ruledtabular}
\end{table}

\begin{figure}[tb]
\includegraphics[width=8.6 cm]{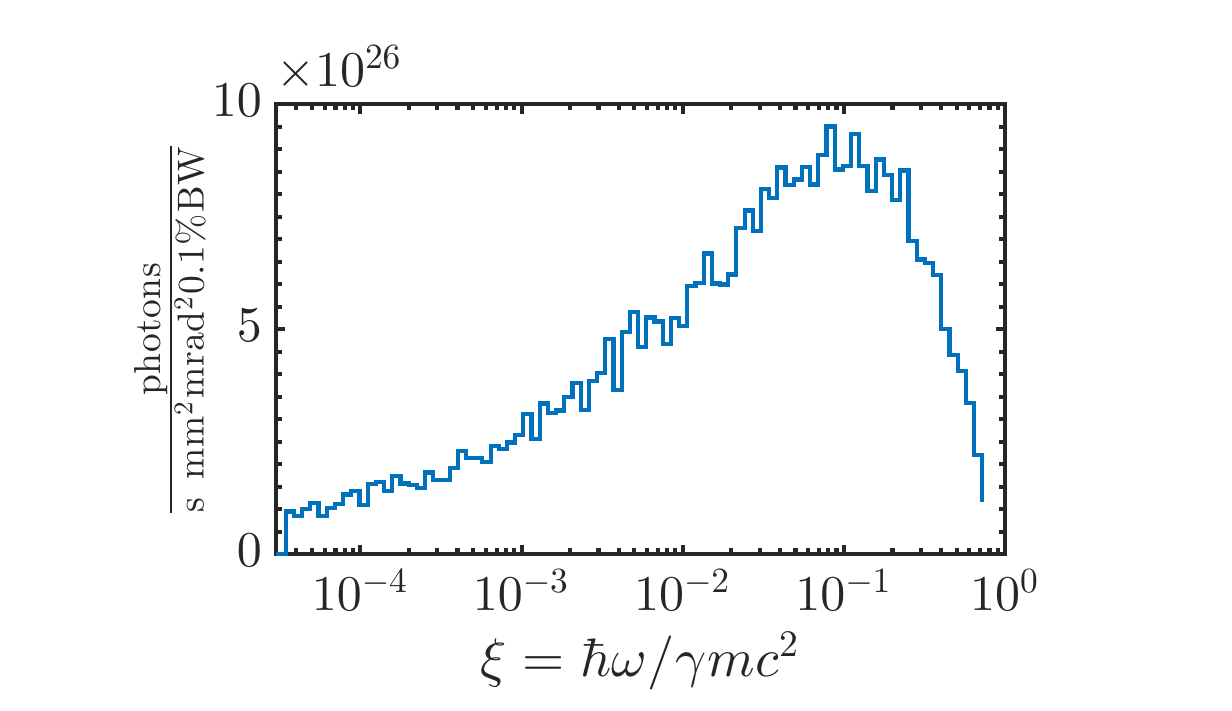}
\caption{ Brilliance of the beamstrahlung $\gamma$ rays source, for two colliding beams of $N=5.8\times10^{10}$ particles, $\mathcal{E}=30$ GeV energy, $\sigma_0=1~\mathrm{\mu m}$ spot size and $\sigma_z=3~\mathrm{\mu m}$ length.}
\label{fig: Brilliance}
\end{figure}

Compared to free electron laser sources, beamstrahlung radiation sources have a broader spectrum, providing higher energy photons but lower brilliance.
{\color{black} Bremsstrahlung sources from relativistic electron beams shot into a solid target produce a broad spectrum in the kev-10s of Mev range with a much lower brilliance, $ \lesssim10^{20}$ $\mathrm{ph/s~mm^2 mrad^2 0.1\% BW}$, compared to the sources previously considered~\cite{AlbertPPCF2016} .
The possibility of synchrotron emission from solid targets by leveraging on nanowire array targets or on plasma instabilities has also been recently investigated~\cite{MartinezPPCF2018,BenedettiNP2018}.
}
Compared to storage rings for synchrotron sources, the collective field of the beams (100s of MGauss) is much stronger than the one that bending magnets usually provide (10s of kGauss).
This enables to reach much higher photon energies.
The emission angle of beamstrahlung, $r_eN/\sigma_0\gamma$, gives a higher divergence to the source compared to storage rings where the emission angle is $1/\gamma$.
However, the beamstrahlung source provides a higher brilliance compared to conventional synchrotron sources as the larger divergence of the beam is compensated by a higher flux the emitted photons.
Figure~\ref{fig: Brilliance} shows the brilliance as a function of the normalized photon energy $\xi$.

{\color{black} 
The theoretical, numerical and experimental investigation of radiation reaction~\cite{Vranic2016NPJ,ColePRX2018,PoderPRX2018} and of pair production~\cite{Grismayer2017PRE,Grismayer2016PoP,Vranic2017PPCF} at the Schwinger limit with a beam-laser or a laser-laser setup has known a surge of interest in connection with the upcoming laser facilities~\cite{eli,vulcan,apollon}.
A beam-beam configuration has also been proposed for studying nonperturbative QED far beyond the Schwinger limit \cite{VitalyArxiv2018}.
These facilities are expected to open new opportunities on laboratory astrophysics, e.g. to study pair cascades~\cite{BellPRL2008,NerushPRL2011,BashmakovPP2014,JirkaPRE2016}.
However, the experimental observation of a few nonlinear Breit-Wheeler pairs was only achieved in the SLAC experiment E-144 in the collision of an electron beam with an optical laser~\cite{BurkePRL1997}.
In the SLAC experiment E-144 the nonlinear regime for pair production was only approached, $a_0\simeq 0.1 - 1$ (field nonlinearity threshold at $a_0=1$).
For the beam parameters discussed before, $N=5.8\times 10^{10}$ and $\sigma_0=1~\mathrm{\mu m}$ the field nonlinearity parameter for a two colliding beams setup would be $a_0\simeq 41$, thus the nonlinear Berit-Wheeler pair production process could be probed far beyond the threshold $a_0\simeq 1$ with a considerable yield, $10^5$, of secondary pairs.}

In conclusion, our results show that in the low disruption regime, beamstrahlung radiation can effectively be employed as highly collimated $\gamma$ rays source at GeV energy with unprecedented brilliance.
Moreover, given the high yield of nonlinear Breit-Wheeler pairs, this setup may be a viable alternative for studying QED effects, approaching the Schwinger limit, compared to beam-laser setups.

\begin{acknowledgments}
This work was supported by the European Research Council (ERC-2015-AdG Grant 695088), FCT (Portugal) Grants No. SFRH/IF/01780/2013 and No. PD/BD/114323/ 2016 in the framework of the Advanced Program in Plasma Science and Engineering (APPLAuSE, FCT Grant No. PD/ 00505/2012). and by the U.S. National Science Foundation Grants No. 1500630 and No. ACI-1339893 and U.S. Department of Energy Grant No. DE-SC0014260. Simulations were performed at IST cluster (Portugal). We acknowledge PRACE for awarding us access to MareNostrum at Barcelona Supercomputing Center (BSC), Spain.
\end{acknowledgments}

\appendix

\section{Derivation of the photon spectrum \label{App: Photon}}
In a constant crossed field, the differential probability rate for photon emission is \cite{Ritus1985JSLR}
\begin{equation}
W_{\omega}=\frac{\alpha}{\sqrt{3}\pi\tau_c\gamma} \left[ \int_b^{\infty}K_{5/3}(x)dx+\frac{\xi^2}{1-\xi}K_{2/3}(b) \right], \label{eq: diff prob phot}
\end{equation}
where $b=2\xi/3\chi_e(1-\xi)$ and $\xi = \hbar\omega/\gamma mc^2$.
The modified Bessel function of the second kind is indicated by $K_{\nu}$, the fine structure constant is $\alpha$, and the Compton time is $\tau_c$.
The classical synchrotron spectrum is given by the integral $\int_b^{\infty}K_{5/3}(x)dx$, which can be approximated by $\int_b^{\infty}K_{5/3}(x)dx\simeq \frac{9}{5}b^{-2/3}\exp{(-b)}$.
The second term contributes mainly to the high energy part of the spectrum and can be approximated, in the limit for large arguments, by $K_{2/3}(b)\simeq\sqrt{\frac{\pi}{2b}}\exp{(-b)}$.
The parameter $b$ is a function of time, of space, and of energy, given by $b = \frac{2}{3} b_0 \frac{\xi}{1-\xi} \frac{x\exp{(u^2)}}{1-\exp{(-x^2)}} = \frac{2}{3} b_0 b_{\xi} b_x b_u $, where $x = r/\sqrt{2}\sigma_0$ is the normalized radius and $u=\sqrt{2}ct/\sigma_z$ is the normalized time that accounts for the relative velocity between the two colliding beams.
The constant $b_0$ is $b_0 = \frac{E_s}{4\sqrt{2}\pi e \gamma n_b \sigma_0}$.
The integration of $W_{\omega}$ in time yields
\begin{eqnarray}
s(\xi,x) & = & \int W_{\omega} dt \nonumber \\
& = & \frac{1}{\sqrt{6\pi}}\frac{\alpha\sigma_z}{c\tau_c\gamma} \left[\frac{9}{5}F_{2/3}+\frac{\xi^2}{1-\xi}\sqrt{\frac{\pi}{2}}F_{1/2}\right] .
\end{eqnarray}
We have used the expansion $b_u = \exp{(u^2)}\simeq (1+u^2)$ around the peak of the collective field at $u=0$, and we introduced the function $F_{\nu}=\left(\frac{2}{3} b_0 b_{\xi} b_x\right)^{-\nu}\left(\nu+\frac{2}{3} b_0 b_{\xi} b_x\right)^{-1/2}\exp{\left(-\frac{2}{3} b_0 b_{\xi} b_x\right)}$.
The weighted integration over space of $F_{\nu}$ is $\int q(x) F_{\nu}(x,\xi) dx$, with $q(x)=2x\exp{(-x^2)}$ the transverse density profile of the beam.
For the integration, we approximate the function $F_{\nu}(x,\xi)$ with a Gaussian form $f_{\nu}(x,\xi)= l_{\nu}(\xi)\exp{[-m_{\nu}(\xi)(x-\hat{x})^2]}$. 
The position of the maximum for $F_{\nu}(x,\xi)$ is located at $\hat{x}\simeq 1$.
The second derivative of $F_{\nu}(x,\xi)$ is evaluated at $\hat{x}$ as
\begin{eqnarray}
\partial_x^2F_{\nu}\Big|_{\hat{x}} &=&  - F_{\nu}\left[1+\frac{\nu}{\left(\frac{2}{3} b_0 b_{\xi} b_x\right)}+\frac{1}{2[\nu+\left(\frac{2}{3} b_0 b_{\xi} b_x\right)]} \right] \times \nonumber \\ 
& & \left(\frac{2}{3} b_0 b_{\xi}\right)\partial_x^2b_x\Big|_{\hat{x}} \nonumber \\
& = & - \left[1+\frac{\nu}{\hat{b}}+\frac{1}{2(\nu+\hat{b})}\right] \frac{2(2\hat{x}-1) \hat{b}F_{\nu}(\hat{b})}{\exp{(\hat{x}^2)}-1} ,
\end{eqnarray}
where $\hat{b} = \frac{2}{3} b_0 b_{\xi}b_x(\hat{x})$.
The parameters $l_{\nu}(\xi)$ and $m_{\nu}(\xi)$ are then obtained as 
\begin{eqnarray}
l_{\nu}(\xi) & = & F_{\nu}(\hat{x},\xi) ,\\
 m_{\nu}(\xi) & = & -\frac{\partial_x^2F_{\nu}(x,\xi)\Big|_{\hat{x}}}{2l_{\nu}(\xi)} \nonumber \\
 & = & \left[1+\frac{\nu}{\hat{b}}+\frac{1}{2(\nu+\hat{b})}\right]\frac{(2\hat{x}-1) \hat{b}}{\exp{(\hat{x}^2)}-1} .
\end{eqnarray}
The weighted integration over space $\int q(x) F_{\nu}(x,\xi) dx\simeq\int q(x) f_{\nu}(x,\xi) dx=\Xi_{\nu}(\xi)$ then gives
\begin{eqnarray}
\Xi_{\nu}  & = & \frac{l_{\nu}\exp{(-\hat{x}^2m_{\nu})}}{1+m_{\nu}} \left\lbrace 1+\sqrt{\pi}\frac{\hat{x}m_{\nu}}{\sqrt{1+m_{\nu}}} \times \right. \nonumber \\
& & \left. \exp{\left(\frac{\hat{x}^2m_{\nu}^2}{1+m_{\nu}}\right)} \left[ 1+\mathrm{erf}\left(\frac{\hat{x}m_{\nu}}{\sqrt{1+m_{\nu}}}\right) \right] \right\rbrace , \label{eq. Xi}
\end{eqnarray}
and the photon spectrum reads
\begin{eqnarray}
\mathcal{S}_{\omega}(\xi)  & = & \int q(x) s(x,\xi) dx \nonumber \\
& \simeq &  \frac{1}{\sqrt{6\pi}}\frac{\alpha\sigma_z}{c\tau_c\gamma} \left[\frac{9}{5}\Xi_{2/3}+\frac{\xi^2}{1-\xi}\sqrt{\frac{\pi}{2}}\Xi_{1/2}\right] .
\end{eqnarray}

\section{Derivation of the yield of secondary pairs \label{App: Pairs}}
In a constant crossed field, the rate of pair creation can be computed as \cite{Ritus1985JSLR}
\begin{equation}
W_p \simeq \frac{2\pi\alpha}{25\tau_c\gamma}\frac{1}{\psi}\exp{\left(-\frac{2\psi}{\xi}\right)} ,
\end{equation}
valid in the limit $\chi_e\lesssim 1$, with $\psi=4/3\chi_e= \frac{4}{3}b_0 b_x b_u$.
The probability of creating a pair from a photon with energy $\xi$ emitted at time $t$ is
\begin{eqnarray}
\mathcal{P}_{\omega\rightarrow p} & = & \int_t^{\infty}W_p dt' \nonumber \\
 & = & \frac{\pi\alpha\sigma_z}{25c\tau_c\gamma}\sqrt{\frac{\pi}{2}}\frac{3\exp{\left(-\frac{8b_0 b_x}{3\xi}\right)}}{4b_0 b_x\sqrt{1+8b_0 b_x/3\xi}} \times \nonumber \\
 & & \left[1-\mathrm{erf}(u\sqrt{1+8b_0 b_x/3\xi})\right] .
\end{eqnarray}
We employed the Taylor expansion $b_u = \exp{(u^2)}\simeq (1+u^2)$ around the peak of the collective field at $u=0$.
The joint probability to produce a secondary pair from a primary particle is then $y_p(x) = \int \int W_{\omega}\mathcal{P}_{\omega\rightarrow p} dt d\xi=\int s(x,\xi)\mathcal{P}_{\omega\rightarrow p}(x,\xi,u=0) d\xi$. By using the exponential function $\exp{\left[-\frac{8}{3\xi}b_0 b_x-\frac{2}{3} b_0 b_{\xi} b_x\right]}=\exp{\left[-\frac{2b_0b_x}{3}\left(\frac{4}{\xi}+\frac{\xi}{1-\xi}\right)\right]}$ with the maximum located at $\hat{\xi}=2/3$, the saddle point method gives
\begin{equation}
y_p(x) = s(x,\hat{\xi})\mathcal{P}_{\omega\rightarrow p}(x,\hat{\xi},u=0)  \sqrt{\frac{\pi}{27b_0b_x}} .
\end{equation}
By denoting $\psi_x = \frac{4}{3}b_0 b_x=\frac{2}{3} b_0 b_{\xi}(\hat{\xi}) b_x$, $y_p(x) $ is
\begin{eqnarray}
y_p(x) & = & \frac{1}{\sqrt{6}}\left(\frac{\pi\alpha\sigma_z}{15c\tau_c\gamma}\right)^2 \left[\frac{9}{5}\sqrt{\frac{2}{\pi}}\frac{\psi_x^{-2/3}}{\sqrt{2/3+\psi_x}} + \right. \nonumber \\
& & \left. \frac{4}{3}\frac{\psi_x^{-1/2}}{\sqrt{1/2+\psi_x}}\right]\frac{\psi_x^{-3/2}\exp{\left(-4\psi_x\right)}}{\sqrt{1+3\psi_x}} .
\end{eqnarray}
We define the function $ F_{\nu} =\frac{\psi_x^{-(3/2+\nu)}\exp{\left(-4\psi_x\right)}}{\sqrt{(\nu+\psi_x)(1+3\psi_x)}}$, analogously to what used for the photon spectrum.
The weighted integration over space of $F_{\nu}$ is $\int q(x) F_{\nu} dx$, with $q(x)=2x\exp{(-x^2)}$ the transverse density profile of the beam.
For the integration, we approximate the function $F_{\nu}$ with a Gaussian form $f_{\nu}= l_{\nu}\exp{[-m_{\nu}(x-\hat{x})^2]}$. 
The position of the maximum of $F_{\nu}$ is also located at $\hat{x}\simeq1$.
The second derivative of $F_{\nu}$ is evaluated at $\hat{x}$ as
\begin{eqnarray}
\partial_x^2F_{\nu}\Big|_{\hat{x}} & = & -F_{\nu}(\hat{x})\left[4+\frac{3/2+\nu}{\hat{\psi}} + \right. \nonumber \\
& & \left. \frac{1}{2}\frac{(1+3\hat{\psi})+3(\nu+\hat{\psi})}{(\nu+\hat{\psi})(1+3\hat{\psi})} \right] \frac{2(\hat{x}-1)\hat{\psi}}{\exp{(\hat{x}^2)}-1} ,
\end{eqnarray}
where $\hat{\psi} = \frac{4}{3} b_0 b_x(\hat{x})$.
The parameters $l_{\nu}$ and $m_{\nu}$ are then obtained as 
\begin{eqnarray}
l_{\nu} & = & F_{\nu}(\hat{x}) ,\\
 m_{\nu} & = & -\frac{\partial_x^2F_{\nu}(x)\Big|_{\hat{x}}}{2l_{\nu}} = \left[4+\frac{3/2+\nu}{\hat{\psi}} + \right. \nonumber \\
 & & \left. \frac{1}{2}\frac{(1+3\hat{\psi})+3(\nu+\hat{\psi})}{(\nu+\hat{\psi})(1+3\hat{\psi})} \right] \frac{2(\hat{x}-1)\hat{\psi}}{\exp{(\hat{x}^2)}-1} .
\end{eqnarray}
The weighted integration over space $\int q(x) F_{\nu}(x) dx\simeq\int q(x) f_{\nu}(x) dx=\Xi_{\nu}$ gives the same function in $l_{\nu}$ and $m_{\nu}$ obtained previously, eq.(\ref{eq. Xi}).
The yield of secondary pairs then reads
\begin{equation}
\mathcal{Y}_p \simeq \frac{1}{\sqrt{6}}\left(\frac{\pi\alpha\sigma_z}{15c\tau_c\gamma}\right)^2 \left[ \frac{9}{5} \sqrt{\frac{2}{\pi}} \Xi_{2/3} + \frac{4}{3} \Xi_{1/2}  \right] .
\end{equation}

\section{The local constant crossed field approximation \label{App: LCFA}}
In the case of an arbitrary field the differential probability rates for photon emission $W_{\omega}$ and for pair production $W_p$ depend on three Lorentz invariants: $f$, $g$, and $\chi$, defined below.
For a constant crossed field ($E = B$, and ${\bf E}\cdot{\bf B}=0$), the invariants $f$ and $g$ vanish.
In the general case, according to Ritus\cite{Ritus1985JSLR}, if the fields are much smaller than $E_s$ then $f,g \ll 1$.
Moreover, if the particle have a large Lorentz factor $\gamma  \gg 1$, these two invariants are also much smaller than $\chi^2$.
Under these conditions, $f$ and $g$ can be neglected and the probabilities $W_{\omega,p}(f,g,\chi)$ are well approximated by $W_{\omega,p}(0,0,\chi)$.
In the proper frame of reference of the beam (primed) there is only an electrostatic field ${\bf E}'$, thus the invariant 
\begin{eqnarray}
g=\frac{F'^{\mu,\nu}F'^{*}_{\mu,\nu}}{4E_s^2}=\frac{{\bf E}'\cdot {\bf B}'}{E_s^2}
\end{eqnarray}
is identically zero.
The difference in amplitude between ${\bf E}'$ and ${\bf B}'$ is determined by the invariant 
\begin{eqnarray}
f &=&\frac{F'^{\mu,\nu}F'_{\mu,\nu}}{4E_s^2} \\
&=&\frac{E'^2-B'^2}{E_s^2}=\frac{E'^2}{E_s^2}.
\end{eqnarray}
This invariant relates to $\chi^2$ of a particle with velocity ${\bf v}'$ and Lorentz factor $\gamma'$ as $f= \chi^2(E'^2_{\perp}+E'^2_{\parallel})/(\gamma'^2E'^2_{\perp}+E'^2_{\parallel})$, where the parallel and perpendicular components are defined with respect to the velocity.
Thus, as the ordering is $E'_{\perp}\sim E'_{\parallel}\ll\gamma'E'_{\perp}$ then $f\sim\chi^2/\gamma'^2\ll\chi^2$, which shows that $f,g \ll 1$ and $f,g \ll \chi^2$.

\begin{figure}[tb]
\includegraphics[width=8.6 cm]{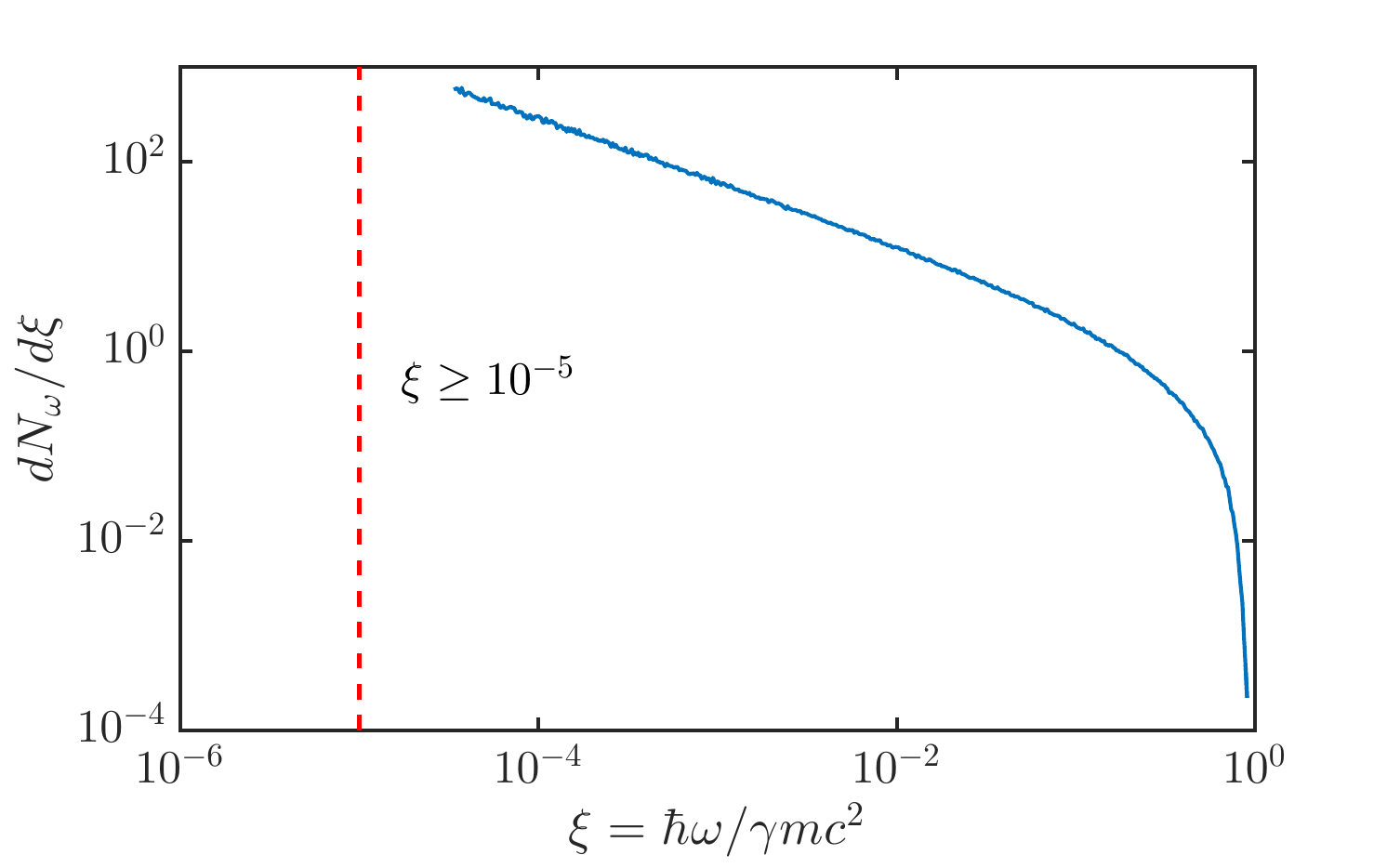}
\caption{Photon spectrum (solid blue) for beam parameters of $N=5.8\times10^{10}$, $\mathcal{E}=30$ GeV, $\sigma_0=1$ $\mu m$ and $\sigma_z=3$ $\mu m$. The local constant field approximation limit of validity (dashed red) is respected by all photons in the simulation.}
\label{fig: LCFA_lim}
\end{figure}

Finally, if the formation length $L_f=mc^2/eE_{\perp}$ is much smaller than the gradient scale of the field $|E_{\perp}/\nabla E_{\perp}|\sim\sigma_z$ the emission process is local and the field leading to the emission can be considered as constant. This condition can be cast in the form $\sigma_0\sigma_z\gg d_e^2$ which breaks for diluted beams but holds for dense beams where the electron skin depth $d_e$ is much smaller than the beam dimensions.
In the conditions explored in the paper, $\sigma_0\sim 5 d_e$, $\sigma_z=3\sigma_0$ and thus $\sigma_0\sigma_z\sim75d_e^2\gg d_e^2 $.
For the above mentioned reasons, the collective field of a dense relativistic beam can be regarded as a constant crossed field, where the differential probability rates for photon emission an for pair production derived by Ritus \cite{Ritus1985JSLR} are applicable.
However, the local constant field approximation might break for the description of nonlinear Compton scattered photons with energy lower than $\xi\ll \chi/a_0^3$~\cite{DiPiazzaPRA2018}.
Figure \ref{fig: LCFA_lim} shows that all photons in our simulation do respect the local constant field approximation limit of validity.
Here the equivalent $a_0$ of the beam is $a_0=41$ and the worst case scenario is at $\chi=1$, this determines that all simulated photons must respect $\xi\geq 1.45\times 10^{-5}$.


\end{document}